\documentclass[12pt]{article}
\usepackage{amssymb,amsfonts}
\def \la{\lambda}

\def \si{\sigma}
\def \di{\displaystyle}

\newcounter{map}
\newcounter{fel}
\begin{document}
\baselineskip 18pt

\title{Derivation of $R$-matrix from local Hamiltonian density}
\author{P.~N.~Bibikov}

\maketitle

\vskip5mm

\begin{abstract}
A computer algebra algoritm for solving the quantum 
Yang-Baxter equation is presented.
It is based on the Taylor expansion of $R$-matrix which is developed
up to the order $\la^6$.
As an example the classification of $4\times4$ $R$-matrices 
is given.
\end{abstract}

\begin{section}
{Introduction}
\end{section}
By the Yang-Baxter equation (in the braid group form) is called the following
equation \cite{1}:
\begin{equation}
R_{12}(\la-\mu)R_{23}(\la)R_{12}({\mu})=R_{23}(\mu)R_{12}({\la})R_{23}(\la-\mu),
\end{equation}
on the $N^2\times N^2$ matrix $R(\la)$ called $R$-matrix. Here for an 
arbitrary $N^2\times N^2$ matrix $X$ the $N^3\times N^3$ matrices
$X_{12}$ and $X_{23}$ are defined as 
$X_{12}=X\otimes I_N$, $X_{23}=I_N\otimes X$,
where $I_N$ is a unit $N\times N$ matrix.

In this paper we study only the regular solutions of the Eq. (1).
These solutions play an important role in the Quantum Inverse Scattering
Method \cite{2} and additionally satisfy the following condition:
\begin{equation}
R(0)=C\cdot I_{N^2},
\end{equation}
where $C$ is a numerical constant.

The system (1),(2) is invariant under multiplication of $R(\la)$ on
the arbitrary regular function $f(\la)$ satisying the initial condition
$f(0)=1$:
\begin{equation}
R(\la)\rightarrow f(\la)R(\la).
\end{equation}
More complicated are the $SL(N,{\mathbb C})$ symmetry \cite{1}:
\begin{equation}
R(\la)\rightarrow M^{-1}\otimes M^{-1} R(\la)M\otimes M,\qquad
M\in SL(N,{\mathbb C}),
\end{equation}
and two discrete symmetries:
\begin{eqnarray}
R(\la)\rightarrow R^t(\la),\\
R(\la)\rightarrow PR(\la)P.
\end{eqnarray}
Here $P$ is the $N^2\times N^2$ exchange matrix defined as operator
in ${\mathbb C}^N\otimes{\mathbb C}^N$ by the formula:
\begin{equation}
P_{ij,kl}=\delta_{il}\delta_{jk}.
\end{equation}
 
Our method of solving the Eq. (1) is based on the Taylor expansion
of $R$-matrix:
\begin{equation}
R(\la)=\sum_{n=0}^{\infty}\frac{1}{n!}R^{(n)}\la^n.
\end{equation}
If we take for simplicity in the Eq. (2) $C=1$ then it follows that $R^{(0)}=I$. 
(From now we shall write $I$ instead of
$I_{N^2}$). The matrix $R^{(1)}$ plays a key role in our approach.
As the local Hamiltonian density
of the integrable model corresponding to $R$ \cite{2}
it is usually denoted by $H$. 
Taking in mind applications to the Inverse Scattering Method we shall 
study only nontrivial Hamiltonian densities. These ones can not
be represented in the form $H=I_N\otimes H_1+H_2\otimes I_N$, with
$N\times N$ matrices $H_1$ and $H_2$.

Matrices $R^{(n)}$ satisfy an infinite system of equations
equivalent to the Eq. (1). These equations does not depend on the
parameter $\la$ and can be consequently solved using the Buchberger
algoritm \cite{3}. 

In \cite{4} this algoritm was succesfully applied to the solution 
of the constant
Yang-Baxter equation with the $R$-matrix does not depending on $\la$.
(Of course the condition (2) was not necessary in \cite{4}). 

Our method of solving the Yang-Baxter equation is based on the well known
fact that for all $R$-matrices which were found up to now the quotents
of their matrix elements may be rationally expressed on $\la$,
${\rm e}^{\la}$, or Jacoby elliptic functions on $\la$. So studying
the first terms of Taylor expansions for these quotents we may try to find
them whole.

The plan of the paper is the following. In the Sect. 2 we study the
expansion (8) up to the order $n=6$. In the Sect. 3 using the $SL(2)$
symmetry (4) we separate all $4\times4$ 
Hamiltonian densities into standard classes.
In the Sect. 4 using the equations presented in the Sect. 2 we find
the complete list of integrable $4\times4$ Hamiltonian densities.
In the Sect. 5 we obtain the corresponding $R$-matrices.

\begin{section}
{The Taylor expansion for $R$-matrix}
\end{section}
As it was mentoined in the previous section the first two terms of the Taylor
expansion are:
\begin{equation}
R^{(0)}=I,\qquad R^{(1)}=H.
\end{equation}

There are two ways to obtain the higher terms of the 
Taylor expansion of $R$-matrix.
First we may do it directly by substitution of partial sums into the Eq. (1).
Another way is to differentiate the Eq. (1) by $\la$ and $\mu$ and then
to take the limit $\la\rightarrow0$, $\mu\rightarrow0$. Technically the second
method is more compact.
However the first method gives a more detailed picture 
especially for $n\geq6$ when in each order of $\la$ there are more than one
equation.

In the second order we obtain the following relation:
\begin{equation}
R^{(2)}_{23}-R^{(2)}_{12}=H^2_{23}-H^2_{12}.
\end{equation}
Its solution will be $R^{(2)}=H^2+\gamma I$.
where $\gamma$ is an arbitrary numerical parameter. The term $\gamma I$ may be 
eliminated by the symmetry (3) if the function
$f(\la)$ satisfy the following conditions: $f'(0)=0$ and
$f''(0)=-\gamma$. So we may write the following expression for $R^{(2)}$:
\begin{equation}
R^{(2)}=H^2,
\end{equation}

In the order $\la^3$ we obtain:
\begin{equation}
R^{(3)}_{23}-R^{(3)}_{12}=H^3_{23}-H^3_{12}+[(H_{12}+H_{23}),J],
\end{equation}
where $J=[H_{12},H_{23}]$.
The solvability of the Eq. (12) leads to the following condition:
\begin{equation}
[(H_{12}+H_{23}),J]=K_{23}-K_{12},
\end{equation}
where $K$ is a $N^2\times N^2$ matrix.

The equation (13) has the standard form:
\begin{equation}
B=F_{23}-F_{12},
\end{equation}
with $N^3\times N^3$ matrix $B$ and $N^2\times N^2$ matrix $F$ 
The analogous condition also appears in
the order $\la^5$. For studying the Eq. (14)
let us consider $F$ as an operator in ${\mathbb C}^N\otimes{\mathbb C}^N$
and $B$ as an operator in ${\mathbb C}\otimes{\mathbb C}\otimes{\mathbb C}$.
For $i=1,2,3$ let ${\rm tr}_i$ be traces in the corresponding subspaces.
For example if $M_1$, $M_2$ and $M_3$ are some $N\times N$ matrices then
${\rm tr}_1(M_1\otimes M_2\otimes M_3)={\rm tr}M_1\cdot M_2\otimes M_3$ 
and so on.

If now for a given matrix $B$ the Eq. (14) has a solution then the matrix $F$ 
may be obtained by the following formula:
\begin{equation}
F=\frac{1}{N}{\rm tr}_1B+\frac{1}{N^2}{\rm tr}_1{\rm tr}_2B\otimes I_N+
\gamma I,
\end{equation}
where $\gamma$ is an arbitrary parameter.
This result may be easely proved if we represent the matrix $B$
in the general form:
\begin{equation}
B=\sum_iX_i\otimes Y_i+I_N\otimes Z_i+T_i\otimes I_N+\alpha I_{N^2},
\end{equation}
where ${\rm tr}X_i={\rm tr}Y_i={\rm tr}Z_i={\rm tr}T_i=0$.

Combining Eqs. (14)  and (15) we obtain a system of linear equations on the
elements of the matrix $B$. For the Eq. (13) this approach leads to the
system of cubic equations on the elements of $H$.
We shall solve it using the Buchberger algoritm. 

In the order $\la^3$ we have met the first obstacle for integrability of the
Hamiltonian density. If the matrix $H$ depends 
on some parameters then solving the Eq. (13)
we shall find the first system of restrictions on them.
 
From (12) and (13) follows that:
\begin{equation}
R^{(3)}=H^3+K.
\end{equation}

The higher order expansion gives:
\begin{eqnarray}
R^{(4)}&=&H^4+2(HK+KH),\\
R^{(5)}&=&H^5+L+2(KH^2+H^2K)+6HKH,\\
R^{(6)}&=&H^6+KH^3+H^3K+9(H^2KH+HKH^2)+\nonumber\\
&&\qquad\qquad\qquad\qquad\qquad+10K^2+3(HL+LH).
\end{eqnarray}
The matrix $L$ is defined by the formula:
\begin{eqnarray}
L_{23}-L_{12}&=&[H_{12}^3+H_{23}^3+3(K_{12}+K_{23}),J]+
3(H_{12}[J,H_{12}]H_{12}+
H_{23}[J,H_{23}]H_{23})+\nonumber\\
&&(H_{12}H_{23}+H_{23}H_{12})(K_{23}-K_{12})+(K_{23}-K_{12})(H_{12}H_{23}+
H_{23}H_{12})-\nonumber\\
&&2(H_{12}(K_{23}-K_{12})H_{23}+H_{23}(K_{23}-K_{12})H_{12}).
\end{eqnarray}

The formulas (19) and (21) follow from the relation:
\begin{eqnarray}
R_{23}^{(5)}-R_{12}^{(5)}+2(H_{12}^2R_{23}^{(3)}-R_{12}^{(3)}H_{23}^2)+
3(H_{12}R_{23}^{(4)}-R_{12}^{(4)}H_{23})&=&\nonumber\\
H_{23}(R_{12}R_{23})^{(4)}-(R_{12}R_{23})^{(4)}H_{12},&&
\end{eqnarray}
which may be obtained from the Eq. (1) by differentiating it one time on $\mu$
and four times on $\la$ and taking the limit: $\la\rightarrow0$,
$\mu\rightarrow0$. Here in (22) we denoted by $(R_{12}R_{23})^{(n)}$ the
following binomial sum:
\begin{equation}
(R _{12}R_{23})^{(n)}=\sum_{i=0}^nC_n^kR_{12}^{(i)}R_{23}^{(n-i)},
\end{equation}
where as usual $C_n^k=\frac{\di n!}{\di k!(n-k)!}$.

Using the similar procedure in the order $\la^6$ we have also 
obtained the following equation:
\begin{eqnarray}
R_{23}^{(6)}-R_{12}^{(6)}+4(H_{12}R_{23}^{(5)}-R_{12}^{(5)}H_{23})+
5(H_{12}^2R_{23}^{(4)}-R_{12}^{(4)}H_{23})&=&\nonumber\\
H_{23}(R_{12}R_{23})^{(5)}-(R_{12}R_{23})^{(5)}.&&
\end{eqnarray}

For even $n$ the expressions for $R^{(n)}$ may be easely obtained
from the equation:
\begin{equation}
R_{12}(-\mu)R_{12}(\mu)=R_{23}(\mu)R_{23}(-\mu),
\end{equation}
which follows from the Eq. (1) when $\la=0$.

Let us notice that the Eqs. (20) and (24) are independent. So in the order
$\la^6$ the substitution of the Taylor sum for $R$-matrix into the Eq. (1)
gives two different equations.

\begin{section}
{Classification of $4\times4$ Hamiltonian densities}
\end{section}
In this section we shall analyse the simpliest case of $4\times4$
matrices $H$. In each equivalence class under the action (4) of $SL(2)$
group we shall choose a unique element. 

First let us represent an arbitrary $4\times4$ matrix $H$ as the following
sum:
\begin{equation}
H=\sum_{i,j=1}^3(S_{ij}\si_i\otimes\si_j+
\varepsilon_{ijk}A_{i}\si_j\otimes\si_k+
h^{(s)}_i(I_2\otimes \si_i+\si_i\otimes I_2)+
h^{(a)}_i(I_2\otimes \si_i-\si_i\otimes I_2)).
\end{equation}
Here $S$ is a symmetrical $3\times3$ matrix:
\begin{equation}
S^t=S,
\end{equation}
and $A$, $h^{(s)}$ and $h^{(a)}$ are vectors in ${\mathbb C}^3$. 

Matrices $\si_i$ for $i=1...3$ are the standard Pauli matrices:
\begin{equation}
\si_1=\left(\begin{array}{cc}
0&1\\
1&0
\end{array}\right),\qquad
\si_2=\left(\begin{array}{cc}
0&-i\\
i&0
\end{array}\right),\qquad
\si_3=\left(\begin{array}{cc}
1&0\\
0&-1
\end{array}\right).
\end{equation}

Using the data ($S$,$A$,$h^{(s)}$,$h^{(a)}$) and
the well known fact of local
isomorphism between $SL(2,{\mathbb C})$ and $SO(3,{\mathbb C})$
we may represent the $SL(2)$ action (4) in the following 
equivalent form: 
\begin{eqnarray}
\tilde S&=&{\cal M}S{\cal M}^{-1},\nonumber\\
\tilde A&=&{\cal M}A,\nonumber\\ 
\tilde h^{(s,a)}&=&{\cal M}h^{(s,a)}. 
\end{eqnarray}
Here ${\cal M}$ is the corresponding to $M$ $3\times3$ matrix satisfying
the following equation:
\begin{equation}
{\cal M}^t={\cal M}^{-1}.
\end{equation}

The Eqs. (27) and (30) may be represented in the following form:
\begin{equation}
<Sx,y>=<x,Sy>,\qquad <{\cal M}x,{\cal M}y>=<x,y>,
\end{equation} 
where $<\cdot,\cdot>$ is a bilinear form on ${\mathbb C}^3$ defined as:
\begin{equation}
<x,y>=\sum_{i=1}^3x_iy_i.
\end{equation}
The form $<\cdot,\cdot>$ 
is nondegenerative so that if for some $x\in{\mathbb C}^3$
and every $y\in{\mathbb C}^3$ $<x,y>=0$ then $x=0$.

Our study  of the $SL(2)$ symmetry will be based first of all
on its action on the matrix $S$,
whose standard form will be defined according to the structure
of its eigenspace. Let $k$ be the dimension of the space spanned on the
eigenvectors of $S$. We shall separately consider the following three 
possibilities: $k=3$, $k=2$, and $k=1$ and in each case
define $S_k$ as the standard form of the matrix $S$.

\begin{subsection}
{k=3}
\end{subsection}

In this case we have also to distinguish seven different subcases:
\begin{enumerate}
\item All the eigenvalues are different and nonsero -- the XYZ case.
Then we shall use the following notations: $\nu_1=\nu_x$, $\nu_2=\nu_y$,
$\nu_3=\nu_z$.
\item All the eigenvalues are different and one of them ($\nu_z$) is 
sero -- the XY case.
\item The two eigenvalues coincide ($\nu_x$ and $\nu_y$) and non of the
eigenvalues is zero -- the XXZ case.
\item The two eigenvalues coincide and nonzero, however the third is zero
($\nu_x=\nu_y=\nu$, $\nu_z=0$) -- the XX case.
\item The two eigenvalues are opposite to each other and
the third is zero ($\nu_x=\nu=-\nu_y$, $\nu_z=0$) -- the X(-X) case.
\item The two eigenvalues are zero and the third is nonzero ($\nu_x=\nu_y=0$,
$\nu_z=\nu\neq0$) -- the Z case.
\item All the eigenvalues coincide and nonzero -- the XXX case.
In this case $\nu_x=\nu_y=\nu_z=\nu\neq0$.
\item All the eigenvalues are zero -- the 0 case.
\end{enumerate}

In all of this subcases we may find in the space ${\mathbb C}^3$ the 
eigenbasis of the matrix $S$ which will be orthogonal 
with the correspondence to the form $<\cdot,\cdot>$. When all eigenvalues
of $S$ are different this fact easely follows from the first Eq. (31) and
nondegenerativity of $<\cdot,\cdot>$. 
When some of the eigenvalues are equal to
each other we may use the ortogonalisation procedure. This also
is possible due to the nondegenerativity of $<\cdot,\cdot>$.

In the XYZ and XY cases the orthogonal eigenbasis of $S$ is fixed.
In this basis the matrix $S$ has the following standard form:
\begin{equation}
S_3=\left(\begin{array}{ccc}
\nu_x&0&0\\
0&\nu_y&0\\
0&0&\nu_z
\end{array}\right).
\end{equation}
The Hamiltonian density depends on the nine parameters: $\nu_i$, 
$A_i$, $h^{(s)}_i$ and $h^{(a)}_i$. 

In $XXZ$, $XX$ and $Z$ cases there exist an additional 
symmetry group preserving the matrix $S$. It is generated by the
following blockdiagonal matrices:
\begin{equation}
{\cal M}=\left(\begin{array}{cc}
{\cal M}_0&0\\
0&1
\end{array}\right),
\end{equation}
where the $2\times2$ matrix ${\cal M}_0$ satisfies the relation (30) and
preserves the form:
\begin{equation}
<x,y>_0=\sum_{i=1}^2x_iy_i,
\end{equation}
on ${\mathbb C}^2$. This means that for arbitrary
$x,y\in{\mathbb C}^2$.
\begin{equation}
<{\cal M}_0x,{\cal M}_0y>_0=<x,y>_0.
\end{equation}

Using this symmetry we may reduce $A$ acting on
the 2-vector $A_0=(A_1,A_2)$:
\begin{equation}
A_0\rightarrow{\cal M}_0A_0.
\end{equation}
There are two different possibilities:
\begin{enumerate}
\item $<A_0,A_0>_0\neq0$. Then $A$ may be reduced to $(A_1,0,A_3)$.
\item $<A_0,A_0>=0$. Then $A$ may be reduced to $(A_1,iA_1,A_3)$
\end{enumerate}

In the $XXX$ case the symmetry group is the whole $SL(2)$
and we also have to consider the two possibilities:
\begin{enumerate}
\item $<A,A>\neq0$. In the appropriate basis $A=(0,0,A_3)$.
\item $<A,A>=0$. In the appropriate basis $A=(A_1,iA_1,0)$.
\end{enumerate}

\begin{subsection}
{k=2}
\end{subsection}
Let $\nu_1$ and $\nu_2$ be the eigenvalues of $S$ corresponding to
eigenvectors $e_1$ and $e_2$. We suppose that $\nu_1$ is degenerative.
This means that for some vector $e_0$:
\begin{equation}
Se_0=\nu_1e_0+e_1.
\end{equation}
We may find the triple $(e_0,e_1,e_2)$ satisfying the 
following system of relations:
\begin{eqnarray}
(e_1,e_k)&=&0,\qquad k=1,2,\\
(e_0,e_2)&=&0.
\end{eqnarray}
When $\nu_1\neq\nu_2$ the Eqs. (39) and (40) are satisfied automatically.
The condition $(e_1,e_2)=0$ follows from (31). Also
\begin{equation}
(e_1,e_1)=((S-\nu_1)e_0,e_1)=(e_0,(S-\nu_1)e_1)=0,
\end{equation}
and
\begin{equation}
(e_1,e_2)=((S-\nu_1)e_0,e_2)=(\nu_2-\nu_1)(e_0,e_2)=0.
\end{equation}

For $\nu_1=\nu_2$ the condition (40) is more refined. From the Eq. (39) and
nondegenarativity of $<\cdot,\cdot>$ follows that $(e_0,e_1)\neq0$. So if
the vector $e_2$ does not satisfy the condition (40) we may define a new one
\begin{equation}
\tilde e_2=e_2-\frac{(e_0,e_2)}{(e_0,e_1)}e_1.
\end{equation}
Now all the relations (39),(40) are satisfied for the triple 
$(e_0,e_1,\tilde e_2)$.

The $3\times3$ symmetrical matrix $S$ whose eigenvectors satisfy 
the conditions (39) and (40) has the following canonical representation:
\begin{equation}
S_2=\left(\begin{array}{ccc}
\nu_x+\frac{\di\alpha}{\di 4}&\frac{\di i\alpha}{\di 4}&0\\
\frac{\di i\alpha}{\di 4}&\nu_x-\frac{\di\alpha}{\di 4}&0\\
0&0&\nu_z
\end{array}\right),
\end{equation}
where $\nu_x=\nu_1$, $\nu_z=\nu_2$ and $\frac{\di\alpha}{\di 4}=
\frac{\displaystyle (e_0,e_1)}{\displaystyle (e_0,e_0)}$.

As in the previous case we need to consider now five different subcases:
\begin{enumerate}
\item The XXZ case -- eigenvalues are nonsero and $\nu_x\neq\nu_z$. 
\item The XX case -- $\nu_x=\nu\neq0$, $\nu_z=0$.
\item The Z case -- $\nu_x=0$, $\nu_z=\nu\neq0$.
\item The XXX case -- $\nu_x=\nu_z=\nu$.
\item The 0 case -- $\nu_x=\nu_z=0$.
\end{enumerate}

\begin{subsection}
{k=1}
\end{subsection}
First let us consider the following matrix
\begin{equation}
\tilde S=S-\nu I_4,
\end{equation}
where $\nu$ is the eiqenvalue of $S$. 

The matrix $\tilde S$ is double degenerated. This means that there exist a
triple ($e_{-1}$,$e_0$,$e_1$) with the following properties:
\begin{equation}
\tilde Se_1=0,\qquad\tilde Se_0=e_1,\qquad\tilde Se_{-1}=e_0.
\end{equation}
The matrix $\tilde S^2$ has two different eigenvectors $e_1$ and
$e_0$ with zero eigenvalues. According to (44) it may be reduced to
the following form:
\begin{equation}
\tilde S^2=\beta^2\left(\begin{array}{ccc}
1&i&0\\
i&-1&0\\
0&0&0
\end{array}\right),
\end{equation}
where $\beta$ is a new parameter. The vectors $e_1$, $e_0$ and $e_{-1}$
may be represented as follows:
\begin{equation}
e_1=\beta^2\left(\begin{array}{c}
1\\
i\\
0
\end{array}\right),\qquad
e_0=\left(\begin{array}{c}
0\\
0\\
\rho
\end{array}\right),\qquad
e_{-1}=\left(\begin{array}{c}
1\\
0\\
-\frac{\di\delta}{\di\beta}
\end{array}\right),
\end{equation}
where $\rho$ and $\delta$ are new parameters.

The system (27),(46) defines the explicit form of $\tilde S$:
\begin{equation}
\tilde S=\left(\begin{array}{ccc}
\delta&i\delta&\beta\\ 
i\delta&-\delta&i\beta\\
\beta&i\beta&0
\end{array}\right),
\end{equation}
as well as the condition $\rho=\beta$. Now the matrix $S$ 
may be represented to the following canonical form:
\begin{equation}
S_1=\left(\begin{array}{ccc}
\nu+\delta&i\delta&\beta\\ 
i\delta&\nu-\delta&i\beta\\
\beta&i\beta&\nu
\end{array}\right).
\end{equation}

As in the previous cases we shall consider the two different subcases:
\begin{enumerate}
\item The XXX case $\nu\neq0$.
\item The 0 case $\nu=0$.
\end{enumerate}

\begin{section}
{Classification of integrable Hamiltonian densities}
\end{section}
In this section we present the complete list of 18 classes 
of integrable $4\times4$
Hamiltonian densities. It is a remarkable fact that for obtaining all of 
them it was necessary to use only the Eq. (13). The Eqs. (21) and (24)
have satisfied automatically. We may contend that these solutions really
correspond to integrable matrices $H$ becouse in the next section the 
corresponding $R$-matrices will be presented.

The result was obtained by computer calculation using
the Buchberger algoritm. We have used for this purpose the computer algebra
system MAPLE \Roman{map}, Release 3 \cite{5},\cite{6}. 
The result will be represented as a
list of the following sets:
$G=\{g_1;...;g_n\}$. Each of them is equivalent the solution: $g_1=...=g_n=0$.
In our calculations we have used the command {\sf gsolve()} of the 
Gr$\ddot o$bner package. In order to present solutions in more compact
form we have introduced the following variables: 
$h^{(s,a)}_{\pm}=h^{(s,a)}_1\pm ih^{(s,a)}_2$, 
$A_{\pm}=A_1\pm iA_2$. Sometimes we have used the following notations:
$h^{(s,a)}_z=h^{(s,a)}_3$ and $A_z=A_3$.
Also we have used the standard shortening changing for 
example $A_1;A_2;A_3$ by $A_{1,2,3}$.

As in the previous Section we shall separately consider the three possibilities:
$k=3,2,1$. 

\subsection
{k=3}
\subsubsection{The XYZ case}
\begin{equation}
G_{3,1}=\{A_{1,2,3};h^{(s)}_{1,2,3};h^{(a)}_{1,2,3}\}.
\end{equation}

\subsubsection{The XY case}
\begin{eqnarray}
G_{3,2}=
\{A_{1,2,3};h^{(s)}_{1,2};h^{(a)}_{1,2,3}\}.
\end{eqnarray}

\subsubsection{The XXZ case}
\begin{equation}
G_{3,3}=\{A_{1,2};h^{(s)}_{1,2,3};h^{(a)}_{1,2}\}.
\end{equation}

\subsubsection{The XX case}

\begin{equation}
G_{3,4}=\{A_{1,2};h^{(s)}_{1,2};h^{(a)}_{1,2,3}\}.
\end{equation}

\subsubsection{The X(-X) case}
\begin{equation}
G_{3,5}=\{A_{1,2};h^{(s)}_{1,2,3};h^{(a)}_{1,2,3}\}.
\end{equation}

\subsubsection{The Z case}
Exept the degenerative case of $G_{3,3}$ there are two new solutions:
\begin{eqnarray}
G_{3,6}&=&\{A_{1,2,3};h^{(s)}_{3};h^{(a)}_{1,2,3}\},\\
G_{3,7}&=&\{A_{1,2,3};h^{(s)}_{1,2};h^{(a)}_{1,2}\}.
\end{eqnarray}

\subsubsection{The XXX case}
In this case all the solutions are the degeneratine cases of $G_{3,3}$ and
$G_{2,6}$.

\subsubsection{The 0 case}
Exept the degenerative case of $G_{2,7}$ there is one solution:
\begin{equation}
G_{3,8}=\{A_{1,2};h^{(s)}_{1,2};h^{(a)}_{1,2}\}.
\end{equation}

\begin{subsection}
{k=2}
\end{subsection}
\subsubsection{The XXZ case}
\begin{equation}
G_{2,1}=\{A_{1,2,3};h^{(s)}_{1,2,3};h^{(a)}_{1,2,3}\}.
\end{equation}
\subsubsection{The XX case}
\begin{eqnarray}
G_{2,2\pm}&=&
\{A_{1,2};h^{(s)}_{1,2};h^{(a)}_{1,2};\nu\pm h^{(s)}_3;
2A_3\pm ih^{(a)}_3\},\\
G_{2,3}&=&\{A_{1,2,3};h^{(s)}_{1,2};h^{(a)}_{1,2,3}\}.
\end{eqnarray}
\subsubsection{The Z case}
\begin{eqnarray}
G_{2,4}&=&\{A_{+,3};h^{(s)}_{+,3};h^{(a)}_{1,2};
\alpha\nu+(h^{(s)}_-)^2;\alpha h^{(a)}_3-iA_-h^{(s)}_-\},\\
G_{2,5}&=&\{A_{1,2,3};h^{(s)}_{+,3};h^{(a)}_{1,2,3}\}.
\end{eqnarray}

\subsubsection{The XXX case}
\begin{equation}
G_{2,6}=\{A_{+,3};h^{(s)}_{1,2,3};h^{(a)}_{+,3}\}.
\end{equation}

\subsubsection{The 0 case}
\begin{eqnarray}
G_{2,7}&=&\{A_{+,3};h^{(s)}_{+,3};h^{(a)}_+\},\\
G_{2,8}&=&\{A_+;h^{(s)}_{1,2,3};h^{(a)}_+;h^{(a)}_-A_3-h^{(a)}_3A_-\}.
\end{eqnarray}

\begin{subsection}
{k=1}
\end{subsection}

\subsubsection{The XXX case}
\begin{equation}
G_{1,1}=\{A_{1,2,3};h^{(s)}_{1,2,3};h^{(a)}_{1,2,3}\}.
\end{equation}

\subsubsection{The 0 case}
\begin{equation}
G_{1,2}=\{A_{+,3};h^{(s)}_{+,3};h^{(a)}_+;
\delta h^{(a)}_3+\beta h^{(a)}_--iA_-h^{(s)}_-\}.
\end{equation}

\begin{section}
{The corresponding $R$-matrices}
\end{section}
In this sections we present $R$-matrices corresponding to all 
the solutions presented in the previous sections. 
For some of the $R$-matrices
the parameter $\la$ is as vector: $\la=(\la_1,\la_2,...,\la_n)$.
This happens when the dependence of elements of the $R$-matrix from $\la$
has the following form:
\begin{equation}
R_{i,j}(\la)=R_{ij}(f_1(a_1\la),f_2(a_2\la),...,f_n(a_n\la)).
\end{equation}
If all the parameters $a_i$ are independent we may put $\la_i=a_i\la$
and consider $\la$ as a vector parameter. 

According to their dependence on $\la$ all the presented $R$-matrices
may be separated in four classes: 1. elliptic, 2. trigonometric, 
3. rational-trigonometric, and 4. rational. When to different values
of parameters of Hamiltonian density correspond $R$-matrices of
different classes we have used the additional labeling. The trigonometric
$R$-matrices were labeled by '$t$', the rational-trigonometric by '$tr$', and
the rational by '$r$'.

\begin{subsection}
{k=3}
\end{subsection}

\begin{subsubsection}
{The XYZ case}
\end{subsubsection}

The $R$-matrix corresponding  to $G_{3,1}$ is the Baxter elliptic
solution of the Eq. (1) \cite{1}.

\subsubsection{The XY case}

The $R$-matrix corresponding  to $G_{3,2}$ is the Felderhof elliptic
solution of the Eq. (1) \cite{7}. In the special case it degenerates into
the following trigonometric $R$-matrix:
\begin{equation}
R_{3,2t}=\left(\begin{array}{cccc}
{\rm sh}\eta\,{\rm ch}{\la}+{\rm sh}\la&0&0&{\rm sh}\eta\,{\rm sh}\la\\
0&{\rm sh}\eta\,{\rm ch}\la&{\rm ch}\eta\,{\rm sh}\la&0\\
0&{\rm ch}\eta\,{\rm sh}\la&{\rm sh}\eta\,{\rm ch}\la&0\\
{\rm sh}\eta\,{\rm sh}\la&0&0&{\rm sh}\eta\,{\rm ch}\la-{\rm sh}\la
\end{array}\right),\qquad (h^{(s)}_z)^2=\nu_x\nu_y.
\end{equation}

\subsubsection{The XXZ case}

\begin{eqnarray}
R_{3,3t}&=&\left(\begin{array}{cccc}
{\rm sh}(\la_1+\eta)&0&0&0\\
0&{\rm e}^{\la_2}\,{\rm sh}\eta&\theta\,{\rm sh}\la_1&0\\
0&\frac{\di 1}{\di \theta}\,{\rm sh}\la_1&{\rm e}^{-\la_2}\,{\rm sh}\eta&0\\
0&0&0&{\rm sh}(\la_1+\eta)
\end{array}\right),\\
R_{3,3tr}&=&\left(\begin{array}{cccc}
\la_1+\eta&0&0&0\\
0&\eta{\rm e}^{\la_2}&\theta\la_1&0\\
0&\frac{\di\la_1}{\di\theta}&\eta{\rm e}^{-\la_2}&0\\
0&0&0&\la_1+\eta
\end{array}\right),\qquad\nu_z^2=\nu_x^2+A_z^2.
\end{eqnarray}

\subsubsection{The XX case}

\begin{eqnarray}
R_{3,4t}&=&\left(\begin{array}{cccc}
{\rm sh}(\la_1+\eta)&0&0&0\\
0&{\rm e}^{\la_2}{\rm sh}\eta&\theta\,{\rm sh}\la_1&0\\
0&\frac{\di 1}{\di\theta}\,{\rm sh}\la_1&{\rm e}^{\la_2}\,{\rm sh}\eta&0\\
0&0&0&-{\rm sh}(\la_1-\eta)
\end{array}\right),\\
R_{3,4tr}&=&\left(\begin{array}{cccc}
\la_1+\eta&0&0&0\\
0&\eta{\rm e}^{\la_2}&\theta\la_1&0\\
0&\frac{\di\la_1}{\di\theta}&\eta{\rm e}^{\la_2}&0\\
0&0&0&-\la_1+\eta
\end{array}\right),\qquad (h^{(s)}_z)^2=\nu^2+A_z^2.
\end{eqnarray}

\subsubsection{The X(-X) case}
\begin{equation}
R_{3,5}=\left(\begin{array}{cccc}
{\rm ch}\la_1&0&0&{\rm sin}\la_2\\
0&{\rm cos}\la_2&{\rm sh}\la_1&0\\
0&-{\rm sh}\la_1&{\rm cos}\la_2&0\\
{\rm sin}\la_2&0&0&{\rm ch}\la_1
\end{array}\right).
\end{equation}

\subsubsection{The Z case}

For $h^{(s)}_+h^{(s)}_-\neq0$ the solution $G_{3,6}$ can be
reduced by the $SL(2)$-symmetry to the Felderhof elliptic case. To the case
$h^{(s)}_+h^{(s)}_-=0$ correspond matrices $R_{3,6\pm}$ ($h^{(s)}_{\pm}=0$).
\begin{eqnarray}
R_{3,6+}&=&\left(\begin{array}{cccc}
1&\theta(1-{\rm e}^{\la})&\theta(1-{\rm e}^{\la})&2\theta({\rm ch}\la-1)\\
0&{\rm e}^{\la}&0&\theta(1-{\rm e}^{\la})\\
0&0&{\rm e}^{\la}&\theta(1-{\rm e}^{\la})\\
0&0&0&1
\end{array}\right),\qquad R_{3,6-}=R_{3,6+}^t,\\
R_{3,7}&=&\left(\begin{array}{cccc}
{\rm e}^{\di\la_1 }&0&0&0\\
0&{\rm e}^{\di\la_2}&0&0\\
0&0&{\rm e}^{\di\la_3}&0\\
0&0&0&{\rm e}^{\di\la_4}
\end{array}\right).
\end{eqnarray}

\subsubsection{The 0 case}
\begin{eqnarray}
R_{3,8t\pm}&=&\left(\begin{array}{cccc}
{\rm sh}(\la_1+\eta)&0&0&0\\
0&{\rm sh}\eta{\rm e}^{\la_2}&\mp i{\rm sh}\la_1&0\\
0&\pm i{\rm sh}\la_1&{\rm sh}\eta{\rm e}^{-\la_2}&0\\
0&0&0&-{\rm sh}(\la_1-\eta)
\end{array}\right),\\
R_{3,8tr\pm}&=&\left(\begin{array}{cccc}
\la_1+\eta&0&0&0\\
0&\eta{\rm e}^{\la_2}&\mp i\la_1&0\\
0&\pm i\la_1&\eta{\rm e}^{-\la_2}&0\\
0&0&0&-\la_1+\eta
\end{array}\right),\qquad (h^{(s)}_3)^2=A_3^2.
\end{eqnarray}

\begin{subsection}
{k=2}
\end{subsection}
\subsubsection{The XXZ case}
\baselineskip 24pt
\begin{equation}
R_{2,1}=\left(\begin{array}{cccc}
{\rm sh}(\la+\eta)&0&0&\theta\,{\rm sh}(\la+\eta)\,{\rm sh}\la\\
0&{\rm sh}\eta&{\rm sh}\la&0\\
0&{\rm sh}\la&{\rm sh}\eta&0\\
0&0&0&{\rm sh}(\la+\eta)
\end{array}\right).
\end{equation}
In the special case: $\theta=-4{\rm sh}\eta$ this $R$-matrix was obtained
in \cite{8}.

\subsubsection{The XX case}
\begin{eqnarray}
R_{2,2t\pm}&=&\left(\begin{array}{cccc}
{\rm sh}(\la+\eta)&0&0&\theta\,{\rm sh}\la\\
0&{\rm sh}\eta\,{\rm e}^{-\la}&\pm{\rm e}^{-\eta}{\rm sh}\la&0\\
0&\pm{\rm e}^{\eta}\,{\rm sh}\la&{\rm sh}\eta\,{\rm e}^{\la}&0\\
0&0&0&-{\rm sh}(\la-\eta)
\end{array}\right),\\
R_{2,2r\pm}&=&\left(\begin{array}{cccc}
\la+\eta&0&0&\la\theta\\
0&\eta&\pm\la&0\\
0&\pm\la&\eta&0\\
0&0&0&-\la+\eta
\end{array}\right),\qquad A_3=h^{(a)}_3=0,\\
R_{2,3}&=&\left(\begin{array}{cccc}
{\rm sh}(\la+\eta)&0&0&\theta{\rm sh}2\la\\
0&{\rm sh}\eta&{\rm sh}\la&0\\
0&{\rm sh}\la&{\rm sh}\eta&0\\
0&0&0&-{\rm sh}(\la-\eta)
\end{array}\right).
\end{eqnarray}

\subsection{The Z case}
\begin{eqnarray}
R_{2,4t}&=&\left(\begin{array}{cccc}
{\rm e}^{\la_1}&\theta({\rm e}^{\la_1}-{\rm e}^{-\la_2})&
\theta({\rm e}^{\la_1}-{\rm e}^{\la_2})&
2\theta^2({\rm ch}\la_2-{\rm e}^{\la_1})\\
0&{\rm e}^{-\la_2}&0&\theta({\rm e}^{\la_1}-{\rm e}^{-\la_2})\\
0&0&{\rm e}^{\la_2}&\theta({\rm e}^{\la_1}-{\rm e}^{\la_2})\\
0&0&0&{\rm e}^{\la_1}
\end{array}\right),\\
R_{2,4r}&=&\left(\begin{array}{cccc}
1&\theta\lambda&-\theta\la&\la(\theta^2\la+1)\\
0&1&0&\theta\la\\
0&0&1&-\theta\la\\
0&0&0&1
\end{array}\right),\qquad h^{(s)}_-=0,\\
R_{2,5}&=&\left(\begin{array}{cccc}
{\rm e}^{\la}&\theta\,{\rm e}^{\eta}\,{\rm sh}\la&
\theta\,{\rm e}^{\eta}\,{\rm sh}\la&
\frac{\di \theta^2}{\di2}({\rm ch}2\eta\,{\rm e}^{3\la}+
{\rm sh}2\eta\,{\rm e}^{-\la}-{\rm e}^{\la+2\eta})\\
0&{\rm e}^{-\la}&0&\theta\,{\rm e}^{\eta}\,{\rm sh}\la\\
0&0&{\rm e}^{-\la}&\theta\,{\rm e}^{\eta}\,{\rm sh}\la\\
0&0&0&{\rm e}^{\la}
\end{array}\right).
\end{eqnarray}

\subsubsection{The XXX case}
\begin{equation}
R_{2,6}=\left(\begin{array}{cccc}
\la+\eta&\xi_1\la&-\xi_1\la&
\theta\la+\frac{\di\xi_1\xi_2+\theta}{\di\eta}\la^2\\
0&\eta&\la&\xi_2\la\\
0&\la&\eta&-\xi_2\la\\
0&0&0&\la+\eta
\end{array}\right).
\end{equation}
In the special case: $\xi_1=\xi_2$, $\theta=-\xi_1\xi_2$ this $R$-matrix
was presented in \cite{9}.

\subsubsection{The 0 case}
\begin{eqnarray}
R_{2,7t}&=&\left(\begin{array}{cccc}
1&(\theta+\xi)({\rm e}^{\lambda}-1)&(\theta-\xi)(1-{\rm e}^{-\lambda})&
\Delta\,{\rm sh}\lambda+2(\theta^2+\xi\eta)({\rm ch}\lambda-1)\\
0&{\rm e}^{\lambda}&0&(\theta+\eta)({\rm e}^{\lambda}-1)\\
0&0&{\rm e}^{-\lambda}&(\theta-\eta)(1-{\rm e}^{-\lambda})\\
0&0&0&1
\end{array}\right),\\
R_{2,7r}&=&\left(\begin{array}{cccc}
1&(\theta+\xi)\lambda&(\theta-\xi)\lambda&
\Delta\lambda+(\theta^2+\xi\eta)\lambda^2)\\
0&1&0&(\theta+\eta)\lambda\\
0&0&1&(\theta-\eta)\lambda\\
0&0&0&1
\end{array}\right),\qquad h^{(a)}_3=0,\\
R_{2,8}&=&\left(\begin{array}{cccc}
{\rm ch}\la_1&
\theta\,({\rm e}^{\la_2}-{\rm e}^{-\la_1})&
\theta\,({\rm e}^{-\la_2}-{\rm e}^{\la_1})&
\eta\,{\rm sh}\la_2+2\theta^2({\rm ch}\la_1-{\rm ch}\la_2)\\
0&{\rm e}^{\la_2}&-{\rm sh}\la_1&\theta\,({\rm e}^{\la_1}-{\rm e}^{\la_2})\\
0&{\rm sh}\la_1&{\rm e}^{-\la_2}&\theta\,({\rm e}^{-\la_1}-{\rm e}^{-\la_2})\\
0&0&0&{\rm ch}\la_1
\end{array}\right).
\end{eqnarray}

\begin{subsection}
{k=1}
\end{subsection}

\subsubsection{The XXX case}
\begin{equation}
R_{1,1}=\left(\begin{array}{cccc}
\la+\eta&\theta\la(\la+\eta)&\theta\la(\la+\eta)&
-\frac{\di\theta^2}{\di\eta}\la(\la+\eta)(\la^2+\la\eta+\xi)\\
0&\eta&\la&-\theta\la(\la+\eta)\\
0&\la&\eta&-\theta\la(\la+\eta)\\
0&0&0&\la+\eta
\end{array}\right).
\end{equation}
In the special case $\theta=-1$, $\xi=\eta^2$ this $R$-matrix was
obtained in \cite{8}.

\subsubsection{The 0 case}
In this case we have separately considered the two subcases: 
$\delta h^{(a)}_3\neq0$ and $\delta h^{(a)}_3=0$. In the first
subcase:
\begin{equation}
R_{1,2}=\left(\begin{array}{cccc}
1&\theta\xi_1(1-{\rm e}^{-\la})&\theta\xi_2({\rm e}^{\la}-1)&
2\theta^2(\Delta({\rm ch}\la-1)-{\rm sh}\la)\\
0&{\rm e}^{-\la}&0&\theta\xi_3(1-{\rm e}^{-\la})\\
0&0&{\rm e}^{\la}&\theta\xi_4({\rm e}^{\la}-1)\\
0&0&0&1
\end{array}\right),
\end{equation}
where
\begin{eqnarray}
\xi_1&=&{\rm e}^{-\varphi_1}{\rm ch}\eta+{\rm e}^{\varphi_2}{\rm sh}\eta,
\nonumber\\
\xi_2&=&{\rm e}^{\varphi_1}{\rm ch}\eta+{\rm e}^{-\varphi_2}{\rm sh}\eta,
\nonumber\\
\xi_3&=&{\rm e}^{\varphi_1}{\rm ch}\eta-{\rm e}^{-\varphi_2}{\rm sh}\eta,
\nonumber\\
\xi_4&=&{\rm e}^{\varphi_2}{\rm sh}\eta-{\rm e}^{-\varphi_1}{\rm ch}\eta,
\nonumber\\
\Delta&=&{\rm sh}(\varphi_1+\varphi_2){\rm sh}2\eta.
\end{eqnarray}
In the second subcase:
\begin{eqnarray}
R_{1,2t}&=&\left(\begin{array}{cccc}
1&\theta\xi(1-{\rm e}^{-\la})&\theta\eta({\rm e}^{\la}-1)&
2\theta^2(1-{\rm ch}\la)\\
0&{\rm e}^{-\la}&0&\frac{\di\theta}{\di\xi}({\rm e}^{-\la}-1)\\
0&0&{\rm e}^{\la}&\frac{\di\theta}{\di\eta}(1-{\rm e}^{\la})\\
0&0&0&1
\end{array}\right),\qquad\delta=0,\\
R_{1,2r}&=&\left(\begin{array}{cccc}
1&\theta\xi\la&\theta\eta\la&-\theta^2\la^2+\gamma\la\\
0&1&0&-\frac{\di\theta}{\di\xi}\la\\
0&0&1&-\frac{\di\theta}{\di\eta}\la\\
0&0&0&1
\end{array}\right),\qquad h^{(a)}_3=0.
\end{eqnarray}

Strictly speaking the list of presented $R$-matrices is not full. For example
making the contraction: $\theta\rightarrow\infty$, $\eta\rightarrow
\infty$, $\frac{\di\theta}{\di{\rm e}^{\eta}}\rightarrow\gamma$, we may obtain
from $R_{3,3t}$ the following $R$-matrix:
\begin{equation}
\tilde R_{3,3t}=\left(\begin{array}{cccc}
{\rm e}^{\la_1}&0&0&0\\
0&{\rm e}^{\la_2}&\gamma{\rm e}^{\la_1}&0\\
0&0&{\rm e}^{-\la_2}&0\\
0&0&0&{\rm e}^{\la_1}
\end{array}\right).
\end{equation}
This case corresponds to $\nu_x=iA_z$ and it also may be easely be obtained 
from the following expressions for $\theta$ and $\eta$: 
$\theta=\sqrt{\frac{\di\nu_x+iA_z}{\di\nu_x-iA_z}}$, ${\rm sh}\eta=\sqrt{
\frac{\di\nu_z^2-\nu_x^2-A_z^2}{\di\nu_x^2+A_z^2}}$.

\baselineskip 18pt

\section{Conclusion}
In this paper we have presented a "direct" computer algebra algoritm for 
solving the quantum 
Yang-Baxter equation using the Taylor expansion of $R$-matrix. 
Applying this algoritm in the $4\times4$ case we have obtained a 
complete list of solutions. 

We beleive that our method will be useful for studying of concrete
Hamiltonian densities appearing in various branches of theoretical
physics.

\end{document}